\documentclass[12pt]{iopart}
\usepackage{iopams}  
\usepackage{graphicx}

\begin{document}

\title{Can dynamical synapses produce true self-organized criticality?}

\author{Ariadne de Andrade Costa}
\address{Departamento de F{\'\i}sica, FFCLRP, Universidade de S\~ao Paulo, 14040-901, Ribeir\~ao Preto, SP, Brazil}
\author{Mauro Copelli}
\address{Departamento de F{\'\i}sica, Universidade Federal de Pernambuco, 50670-901, Recife-PE, Brazil}
\author{Osame Kinouchi}
\address{Departamento de F{\'\i}sica, FFCLRP, Universidade de S\~ao Paulo, 14040-901, Ribeir\~ao Preto, SP, Brazil}
\eads{\mailto{ad.andrade.costa@gmail.com}, \mailto{mcopelli@df.ufpe.br}, \mailto{osame@ffclrp.usp.br}}
\begin{abstract}
Neuronal networks can present activity described by power-law distributed avalanches presumed to be a signature of a critical state. Here we study a random-neighbor network of excitable cellular automata coupled by dynamical synapses. The model exhibits a very similar to conservative self-organized criticality (SOC) models behavior even with dissipative bulk dynamics. This occurs because in the stationary regime the model is conservative on average, and, in the thermodynamic limit, the probability distribution for the global branching ratio converges to a delta-function centered at its critical value. So, this non-conservative model pertain to the same universality class of conservative SOC models and contrasts with other dynamical synapses models that present only self-organized quasi-criticality (SOqC). Analytical results show very good agreement with simulations of the model and enable us to study the emergence of SOC as a function of the parametric derivatives of the stationary branching ratio.    
\end{abstract}


\maketitle

\tableofcontents

\section{Introduction}

In the past few decades there has been an increasing exploration of the heuristic idea that the concepts of 
complexity and criticality are intermingled~\cite{Langton90, Wolfram83-RMP, Bak87, Chialvo10, Reia14}.
Criticality is associated with critical $d$-surfaces in parameter space, for example, a critical point ($d=0$-surface) 
or a critical line ($d=1$) in a continuous 
phase transition. If the system has a $D$-dimensional parameter space, to be critical 
at a $d<D$ surface necessarily implies fine tuning of
the system parameters to be close to that surface. Since several complex 
systems seem to lie spontaneously at the border of such
critical surfaces, we need to explain such fine tuning. 

Several mechanisms have been proposed to do that, the most popular being based 
in the notion of Self-Organized Criticality (SOC).
SOC in microscopically conservative systems, whose prototypical example
is the sandpile model~\cite{Bak87}, is by now a well understood issue~\cite{Jensen98, Munoz99, Dickman00}.  
The case of non-conservative (bulk) dynamics has also been subjected to intense 
study and the present consensus is that these systems are critical only in the conservative 
limit~\cite{Drossel92, Socolar93, Broker97, Chabanol97, Jensen98, 
Kinouchi99, deCarvalho00, Bonachela09, Bonachela10, Martin10}. 
In the absence of that, what one observes is
simply a large critical region surrounding the critical point, which can be confounded 
with a generic critical interval due to finite size effects (Pseudo-SOC)~\cite{Broker97, Chabanol97, Jensen98,
Kinouchi99, deCarvalho00}. Another class of models has loading mechanisms that make the 
system hover around the critical point~\cite{Drossel92, Levina07}, with fluctuations
that do not vanish in the thermodynamic limit, a case which has been called Self-Organized
quasi-Criticality (SOqC)~\cite{Bonachela09, Bonachela10}. A review about SOC
applied to neural networks discuss several limitations of the present models~\cite{Hesse2014}.

Here we introduce a new class of models that, although 
non-conservative microscopically, after a transient
achieves a stationary regime which is conservative
on average, with vanishing fluctuations in the thermodynamic limit. 
This new behavior, not seen in previous SOC models, 
is due to an intra-avalanche ultra-soft loading mechanism 
(loading in a slower time scale than that presented in SOqC models). Another advantage of our model
is that it has a very simple, exact and transparent mean-field treatment, in comparison to
other models of the literature.

\section{The model}

First we introduce the Levina-Herrmann-Geisel (LHG) model~\cite{Levina07,Bonachela10}, 
which directly inspired our model. Surprisingly,
we will find that the models pertain to different universality classes.
In the LHG model the basic excitable elements are integrate-and-fire neurons with time dependent membrane
potentials $0<V_i(t)<V_{max}, i=1,\ldots,N$. The synapses are denoted as $J_{ij}(t)$ and also evolve in time.
The synaptic coupling is all-to-all (so there are $N(N-1)$ synapses).
The model dynamics is given by
\begin{eqnarray}
\label{LHG}
\frac{\partial V_i}{\partial t}= I^{ext}\delta(t-t^i_{driv})+\sum_{j=1}^{N-1} \frac{uJ_{ij}}{N-1}\delta(t-t_{sp}^j)-V_{max}\delta(t-t_{sp}^i),\\
\label{LHG_J}
\frac{\partial J_{ij}}{\partial t}= \frac{1}{\tau_J}\left( \frac{\alpha}{u}-J_{ij}\right)-uJ_{ij} \delta(t-t_{sp}^j).
\end{eqnarray}
\begin{itemize}
\item Driving: the $I^{ext}$ term is a slow drive on neuron $i$ that acts on times $t^i_{driv}$ with a given rate.
\item Firing: the neuron spikes if $V_i(t) > V_{max}$, which defines the spike time $t_{sp}^i$. The term $-V_{max}\delta(t-t_{sp}^i)$ corresponds to a reset of the membrane potential.
\item Integration: the sum over $J_{ij}$ corresponds to the integration, by the postsynaptic neuron $i$, 
of the synaptic contributions of all firing presynaptic neurons $j$.
\item Synaptic depression: when a presynaptic neuron $j$ fires (which defines the spike time $t_{sp}^j$), 
all synapses (out-links) $J_{ij}$ are depressed by the amount $uJ_{ij}$.
\item Synaptic recovery: synapses recover to a target value $\alpha/u$ with a time scale given by $\tau_J$.
\end{itemize}

Both LHG~\cite{Levina07} and Bonachela et al.~\cite{Bonachela10} did not study the full parameter space 
$(\alpha,u,\tau_J,V_{max})$ but mainly studied the effect of varying $\alpha$ with other parameters fixed as
$u=0.2, V_{max}=1$ and $\tau_J=10N$. They found a critical-like region around $\alpha = 1.4$. However, Bonachela {\it et al.}
showed that this region is pseudocritical because the system's behavior is an oscillation around the critical point
with an amplitude that does not vanish in the large $N$ limit. 

Our model builds upon a random-neighbor network of excitable neurons used previously~\cite{Kinouchi06a} 
(inspired on the well-known SIRS epidemiological model). The possible states ($S_i =0$, $i=1,\dots, N$) are:
\begin{itemize}
\item Susceptible or Quiescent: $S_i =0$;
\item Infected or Firing: $S_i =1$ ;
\item Recovering or Refractory: $S_i = 2, 3, \ldots, n-1$.
\end{itemize}

The network is of random neighbor type, each presynaptic neuron $j$ having exactly $K$ 
outlinks to postsynaptic neurons $i$  described by probabilistic couplings (synapses) $P_{ij}$. 
Firing sites can induce firing in neighbors sites, creating an avalanche (defined as a not interrupted sequence
of firing sites). The dynamics is composed by the following steps:

\begin{itemize}
\item Driving: After an avalanche occurs and all sites are either quiescent or refractory, we choose a single site at random and
force it to fire ($S_i = 1$), creating a new avalanche. 
\item Firing: the probability that a presynaptic neighbor $j$ does not induce a firing in the postsynaptic neuron $i$ is
$\left[1-P_{ij}\delta(S_j(t),1)\right]$, where henceforth $\delta$ is 
the Kronecker delta function.
So, the probability that a quiescent ($S_i(t)=0$) site spikes ($S_i(t+1)=1$) is given by
$1- \prod_{j=1}^{K_i} \left[1-P_{ij}\delta(S_j(t),1)\right]$, where $K_i$ is the number of incoming links of site 
$i$ (note that $\langle K_i \rangle=K$). 
\item Refractory time: after a site spikes, it deterministically becomes 
refractory for $n-2$ time steps, and then returns to quiescence:

\begin{equation}S_i (t+1) = \left\{\begin{array}{ll}
S_i(t) + 1, \textrm{ if }  S_i = 1,2,\ldots,n-2; \\
0, \textrm{ if } S_i = n-1. \end{array} \right. \end{equation} 

\item Synaptic depression and recovery: to be described below.
\end{itemize}

Each site has a local branching ratio $\sigma_i \equiv \sum_j^K P_{ij}$
and we can define a global branching ratio $\sigma \equiv N^{-1} \sum_i^N \sigma_i$. 
If $\{P_{ij}\}$ are drawn from a uniform distribution with average $\sigma/K$ and 
kept fixed then the average branching ratio $\sigma$ controls the collective 
behavior of the network: for $\sigma < 1$ $(>1)$ the system is subcritical (supercritical), 
with unstable (self-sustained) activity. This is the static version of the model, without dynamical
synapses, which undergoes a continuous 
absorbing state phase transition at $\sigma=\sigma_c=1$~\cite{Kinouchi06a}. For this static model,
criticality is only achieved by fine tuning $\sigma$ to a critical value $\sigma_c$. This absorbing
continuous phase transition pertains to the class of Compact Directed Percolation (CDP) and is 
identical to that present in canonical conservative SOC models \cite{Munoz99, Dickman00, Bonachela09}.

\begin{figure}[!ht]
  \centering  
  \includegraphics[width=0.45\columnwidth]{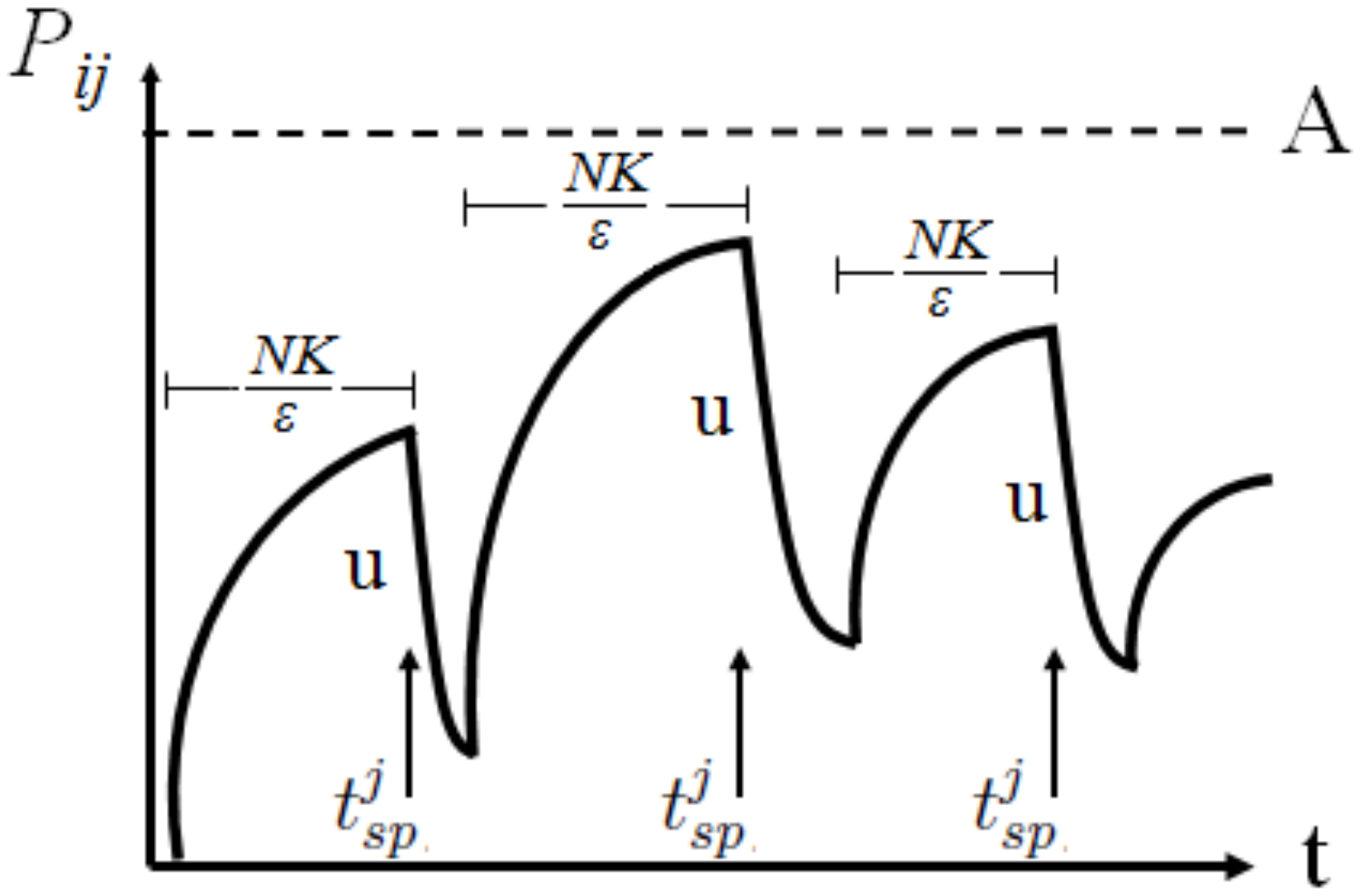}
  \caption{Schematic illustration of Eq.~\ref{Pdyn} parameters. $P_{ij}$ represents the amount of available neurotransmitters at the synapse between the neurons \textit{j} and \textit{i} (in the model, it is the probability of neuron \textit{j} excites neuron \textit{i}). The parameter \textit{A} corresponds to the largest value the synapses can reach; the factor $\epsilon/NK$ is the speed with which synapses recover each time step; and the parameter $u$ is the subtracted fraction of $P_{ij}$ when the presynaptic neuron \textit{j} fires, it is, when $t=t^j_{sp}$.}
   \label{Figmodel}
\end{figure}

Here we report new results obtained with a homeostatic synaptic mechanism described
by time-dependent probabilities $P_{ij}(t)$ that 
follow a depressing/recovering synaptic dynamics similar to that used in the LHG model~\cite{Levina07, Bonachela10}. 
The crucial difference with respect to their model, however, is that the recovering mechanism of our
synaptic dynamics occurs at a slower, $O(1/N)$, time scale:
\begin{equation}
\label{Pdyn}
P_{ij}(t+1)=P_{ij}(t)+\frac{\epsilon}{NK}(A-P_{ij})-uP_{ij}(t) \delta(t,t^j_{sp}) \; ,
\end{equation} 
where $t^j_{sp}$ is the spiking time of the $j$-th presynaptic neuron. 
A comparison with the synaptic part of the LHG model (Eq.~\ref{LHG_J}) gives that $\frac{\epsilon}{NK}$ is proportional to $\frac{1}{\tau_J}$ 
(remember that $\tau_J$ is proportional to $N$) and that $A = \alpha/u$, with $u$ having the same meaning in both models. 
The parameters of Eq.~\ref{Pdyn} can be better understood in Figure~\ref{Figmodel}.

Notice that the dynamics in $P_{ij}(t)$ leads to a time-dependent global branching 
ratio $\sigma(t)=\frac{\sum_{ij} P_{ij}}{NK}$.  
The different time scale from the LHG model arises because 
they use an all-to-all coupling where synapses already have a scale
$1/N$ (like in the all-to-all Ising model). So, synaptic recovery in the LHG model perturbs the synapses
with a term of same order $O(1/N)$  and this could be the origin of the large variance on the average synaptic weight
measured in \cite{Bonachela10}. In our case, $P_{ij}$ is $O(1)$ and the $1/N$ factor in Eq.~(\ref{Pdyn}) defines an
ultra-soft recovery mechanism. Although this could be considered a small modeling change, it produces
an important change in the universality class of our model.

We observe that the scaling $1/N$ in the synapses dynamics, also present in the LHG model, is somewhat problematic from the biological
point of view, since synapses do not have information about how many neurons there are in the network. 
However, in the literature of networks with dynamical links, this feature corresponds to the modeling state of
art and there are no models without this feature. Perhaps, in real networks, the recovery
dynamics mediated by $\epsilon$ does not depend on $N$ but has anyway a very large recovery time
(small $\epsilon$) which is sufficient to put the system in a state very close of the critical one. 
This problem also suggests a research topic: what are the time scales that biological synapses recover after spike and
if the recovery time depends on the size of the network.

We also observe that $\sigma(t)=N^{-1} \sum_{ij}^K P_{ij}(t)$ is a property of the architecture,
and how the network links ${P_{ij}}$ change. It reflects the average number of firings potentially induced
by a firing site in the absence of overall network activity. So, $\sigma$ is not the dynamical branching
ratio measured by averaging the ratio between the number of firing sites over the last firing sites,
by using an actual firing time series, which is equal to one even for a supercritical systems~\cite{Hesse2014}.
The two concepts (branching ratio for the architecture and branching ratio for the activity) have the same 
interpretation only in the critical limit $\sigma \to 1$.

\section{Mean-Field analysis}

The model is amenable to very simple and transparent analytic treatment 
at a mean-field level. We initially follow the steps developed for the static model with fixed
$P_{ij}$~\cite{Kinouchi06a}. Let $\rho_m(t)$ be the density of sites
in state $m \in \{0,1,\ldots,n-1\}$ at time $t$. The ensemble
average of $P_{ij}(t)$ is $\sigma(t)/K$. The probability that a
quiescent neuron at time step $t$ will be excited in the next time
step by at least one of its $K$ neighbors can be
approximated~\cite{Kinouchi06a} by
$ \pi(t)=1-\left[1-\rho(t)\sigma(t)/K\right]^{K} $, 
where $\rho(t) \equiv \rho_1(t)$ is the density of active sites. Given
that a site can only be excited when it is quiescent, the dynamics of
$\rho(t)$ can be written by resorting again to the mean-field
approximation 
$\rho(t+1)=  \pi(t) \rho_{0}(t)$. 
The dynamics of $\rho_{0}(t)$, the density of quiescent neurons, is
coupled to those of the refractory states, whose deterministic
dynamics immediately yield 
in the thermodynamic limit $N\to\infty$: 
\begin{eqnarray}
  \rho_2(t+1) &=& \rho(t)  \\
  \rho_3(t+1) &=& \rho_2(t) \\
  &\vdots & \nonumber \\
  \rho_{n}(t+1) &=& \rho_{n-1}(t) \; .
\end{eqnarray}
To study the stationary state of the mean-field equations, we drop the $t$ dependence in the stationary state ($t\to\infty$) from
the above equations to obtain $\rho_{n-1}^* = \rho_{n-2}^* = \ldots =
\rho_{2}^* = \rho^*$. Imposing the stationary condition onto the
normalization $\sum_{m=0}^{n-1} \rho_m(t)=1$, we arrive at
$\rho_0^* = \left[1-\left(n-1\right)\rho^*\right]$. 
We therefore have, in the stationary state,  
the first of our self-consistent equations, which describes the stationary density
$\rho^*$ of active sites for fixed coupling $\sigma^*$~\cite{Kinouchi06a}:
\begin{equation}
\rho^* = \left[1-\left(n-1\right)\rho^*\right] \left[1-\left(1-\sigma^* \rho^*/K\right)^{K}\right] \; .
\label{rho_fcao_sig}
\end{equation}

By considering the limit $\rho \rightarrow 0$ in
Eq.~(\ref{rho_fcao_sig}), we can obtain the mean field behavior for a
critical branching process: $ \rho^* \simeq 
(n-1)^{-1}\left(\frac{\sigma^*-\sigma_c}{ \sigma_c}\right)^\beta $, 
with the critical value $\sigma_c = 1$ and the usual mean-field
exponent $\beta =1$~\cite{Munoz99}.


In Eq.~(\ref{rho_fcao_sig}),  $\sigma^*$ was
assumed constant. To obtain its value, we impose the stationary
condition on Eq.~(\ref{Pdyn}) for the synaptic dynamics, such that the
dissipation and the driving (loading) of the system must be the
same. Dropping the $t$ dependence and averaging Eq.~(\ref{Pdyn}) over
the ensemble, we obtain
\begin{equation}
\frac{\epsilon}{KN}\left(A-\frac{\sigma^*}{K}\right) = \frac{u\sigma^*\rho^*}{K},
\end{equation}
which can be solved for $\sigma^*$, rendering
\begin{equation}
\sigma^* = \frac{AK\epsilon}{uKN\rho^*+\epsilon} \; .
\label{sigma_estacion}
\end{equation} 
This is the second of our self-consistent equations, stating the
average coupling $\sigma^*$ as a result of the interplay between
synaptic depression ($u$) and recovery ($A$), in light of a constant
density $\rho^*$ of spiking neurons.

Together, Eqs.~(\ref{rho_fcao_sig}) and~(\ref{sigma_estacion}) can be
solved to determine $\rho^*$ and $\sigma^*$. In particular, in the
critical region $\sigma \gtrsim \sigma_c$,  
we can understand how
the model parameters affect the distance of the stationary branching ratio
$\sigma^*$ from what would be the critical value $\sigma_c=1$ in a truly
self-organized system:
\begin{equation}
\label{sigma_dependence}
  \sigma^*-1 \simeq
  \frac{\left(AK-1\right)}{1+x}\; ,  x \equiv \frac{uKN}{\left(n-1\right)\epsilon}
  \; ,
\end{equation} 
where the scaling variable $x$ condenses the effect of most of the parameters and the
important $N$ dependence.  
Therefore, the mean-field calculation predicts that when $x\gg 1$ (which we call large-$N$ 
tuning and is realized for finite $u$, $K$, $\epsilon$ and large $N$), the stationary value
$\sigma^*$ differs from $\sigma_c=1$ by a term of order $1/N$. The several
parameters of the model only affect the constant prefactor of this
term. We have a critical state without the need of fine tuning
of the parameters, requiring only  the large $N$ limit which enables the
evolution of $\sigma(t)$ to approach the region where
Eq.~(\ref{sigma_dependence}) is valid. 

We also note that $\sigma^*=1$ can be produced exactly in our model,
but at the expense of choosing $A=1/K$, which is a fine tuning
operation (pulling $P_{ij}(t)$ towards $P_c = 1/K$).  In short, due to
the synaptic dynamics, $\sigma$ is no longer a parameter (like in
the static model~\cite{Kinouchi06a}) but is rather a slow $\sigma(t)$ dynamical
variable whose stationary value depends on the parameters $\epsilon$,
$A$, $u$, $K$, $n$ and $N$. The coupled equations (\ref{rho_fcao_sig})
and (\ref{sigma_estacion}) are solved numerically to give curves
$\sigma^*(\epsilon,A,u,K,n,N)$ and $\rho^*(\epsilon,A,u,K,n,N)$. 

If we expand Eq.~\ref{rho_fcao_sig} for large $N$ and use Eq.~\ref{sigma_estacion}, we obtain:

\begin{equation}
\label{largeN}
\rho^* \approx \frac{A\epsilon}{uN}.
\end{equation}
This shows that, for large $N$, we have $\rho \rightarrow 0$ (critical state), and $\rho$ grows with 
$(\epsilon, A, 1/u )$ but has a $1/N$ prefactor. So, a graphic of $\rho$ as a function of any parameter
of the model shows that the critical state (and the avalanche behavior) depends very weakly on
$(\epsilon, A, u )$ as $N$ grows. We will call this dependence, which vanishes fast with $N$, as
gross-tuning, to differentiate it from the fine-tuning needed in several models to achieve 
SOC~\cite{Broker97, Chabanol97, Jensen98, Kinouchi99, deCarvalho00, Bonachela09}.  

\section{Results}

Our mean-field calculation describes a system without spacial correlations, in which neighbors are chosen at random
at each time step (annealed model). Although being a step back in biological realism, the annealed model is very
important due to the insights furnished along the mean-field results. 

\begin{figure}[!ht]
  \centering  
  \includegraphics[width=0.45\columnwidth]{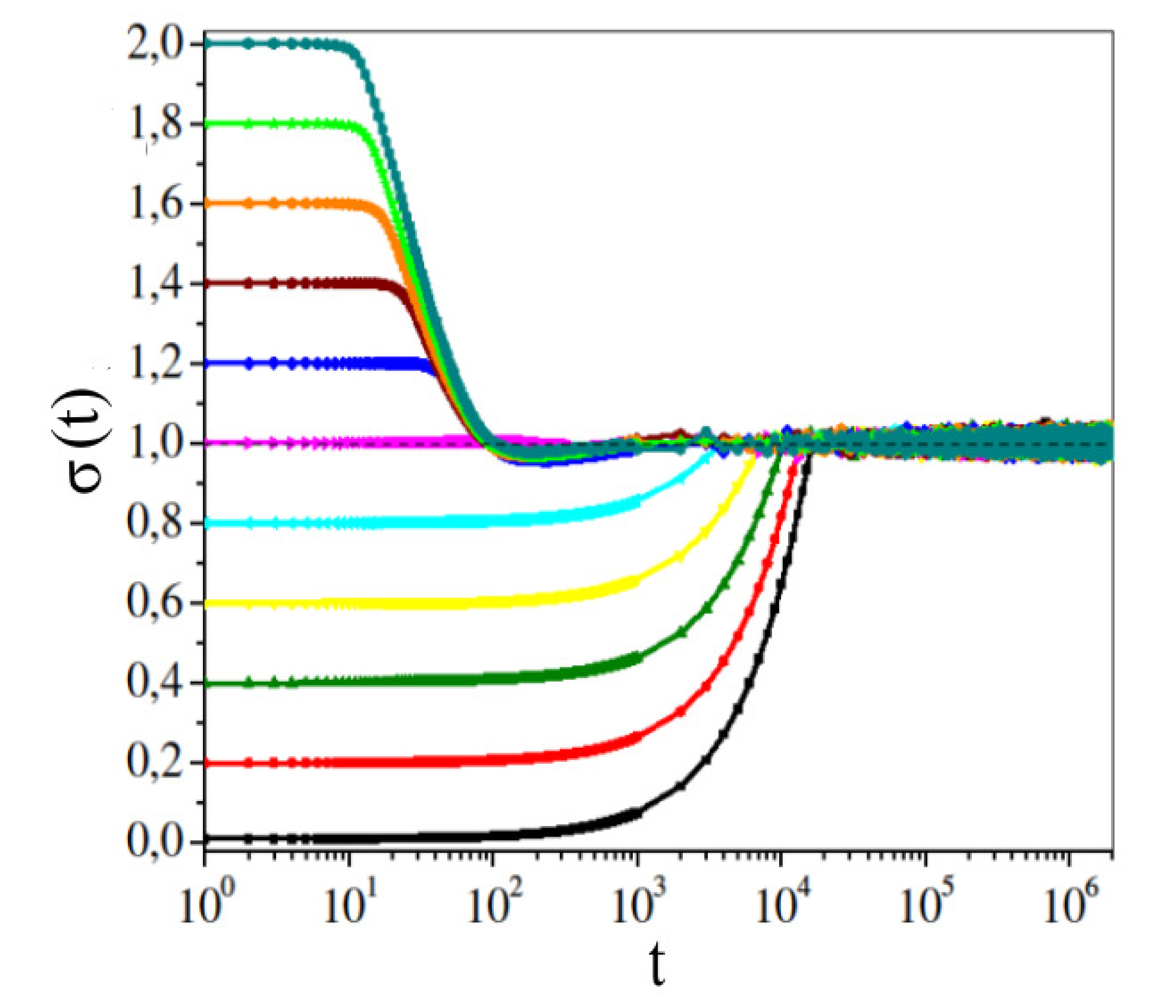}
  \caption{\label{sigma-vs-time} Transient and self-organization for  
  different initial values $\sigma_0$ with parameters: 
$\epsilon=2.0$, $u=0.1$, $A=1.0$, $K=10$, $n=3$ and $N=30000$. 
The dotted line correspond to the stationary branching ratio $\sigma^*$ around which 
the systems oscillates after a transient: $\sigma^*=1.000\pm0.012$. }
\end{figure}

Irrespective of the initial distribution of couplings $P(P_{ij}, t=0)$, which defines an initial
value $\sigma_0$, the network architecture evolves (``self-organizes'') during a transient toward a stationary 
regime $P(P_{ij},t\rightarrow \infty) \equiv P^*(P_{ij})$. 
This self-organization can be followed by measuring $\sigma(t)$, see Fig.~\ref{sigma-vs-time}.
The fact that the branching ratio $\sigma(t)$ evolves and self-organizes in time is a characteristic
of networks with adaptive links not present in classical SOC models like sandpiles and earthquake models, where 
the links are static and represent, say, how much a given toppling site gives to its neighbors. Also, if
perturbed or damaged, the set of synapses recovers and achieves a new stationary state $P^*(P_{ij})$ similar to
the previous one.
The evolution of $P(P_{ij},t)$ and the corresponding $\sigma(t)$ towards criticality, 
as exposed in Fig.~\ref{sigma-vs-time}, seems
for us to be a more strong instantiation of the original idea by Per Bak of a truly self-organizing system~\cite{bak}.

\begin{figure}[!ht]
\centering
 \includegraphics[width=0.6\columnwidth]{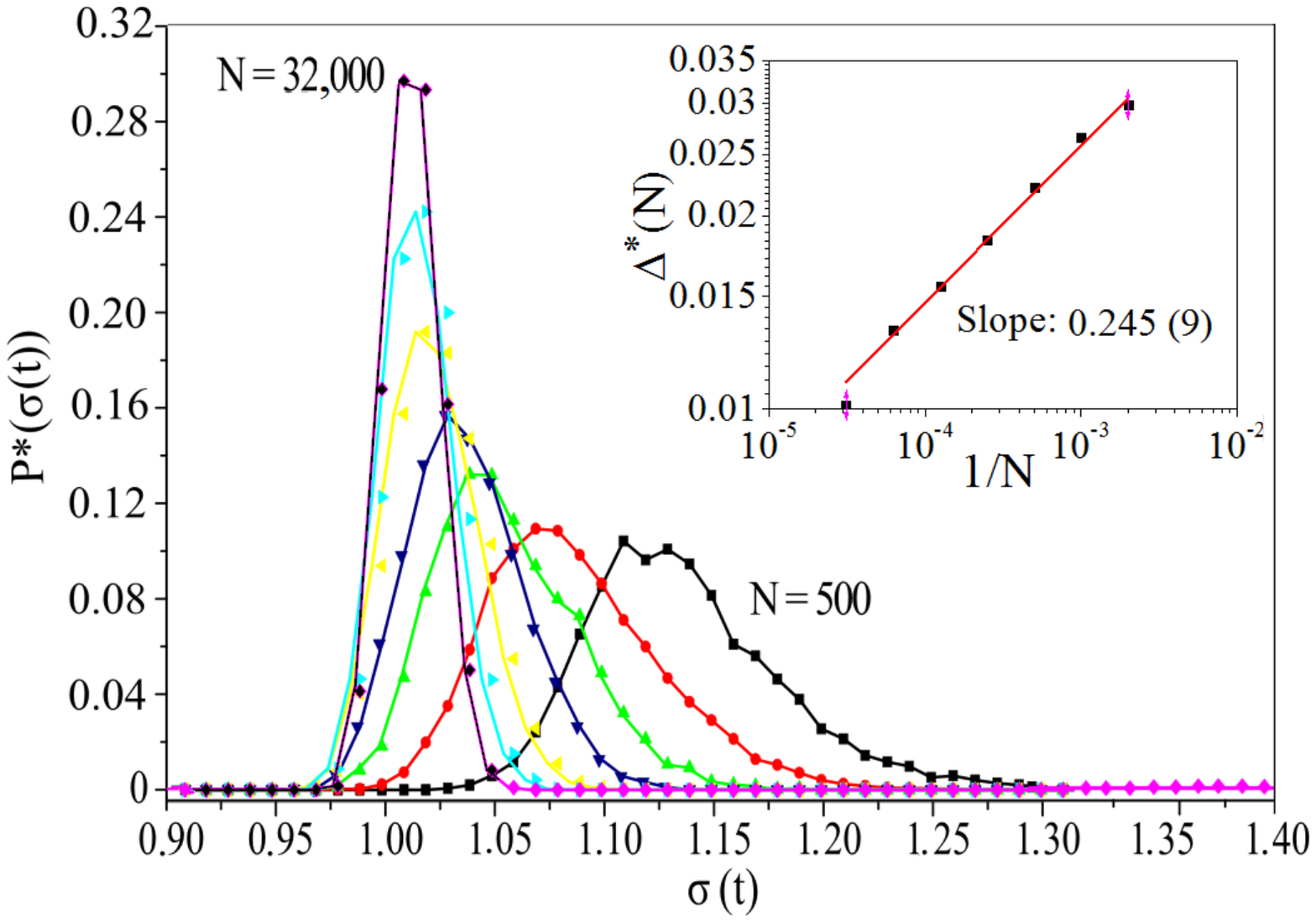}
        \caption{Branching ratio distribution $P^*(\sigma)$ for different network sizes. 
Parameters are: $\epsilon=2.0$, $u=0.1$,
$A=0.9$, $K=10$ and $n=3$. Inset: stationary standard deviation $\Delta^*$ as a function of 
$N = 500$, $1000$, $2000$, $4000$, $\ldots$, $32000$ 
(in double logarithmic scale). Symbols are simulations; line is a least-square fit.}
       \label{P(sig)_vs_sig}
\end{figure}

The stationary time series for $\sigma(t,N)$ presents 
fluctuations around an average value $\sigma^*(N)$, with standard deviation $\Delta^*(N)$. 
The stationary distribution $P^*(\sigma,N)$  is roughly Gaussian 
(Fig.~\ref{P(sig)_vs_sig})  with the large-$N$ scaling $\sigma^*(N)  = \sigma_c + C_\sigma/N$, 
with the critical point $\sigma_c=1$ (see Eq.~(\ref{sigma_dependence})).
The most important result is that it tends to a delta function, 
with $\Delta^*(N) \propto N^{-1/4}$, see inset of Fig.~\ref{P(sig)_vs_sig}.

In the stationary state, the model is therefore conservative on average, 
in the sense that it conserves the average number of active sites. In other words, 
its time-averaged branching ratio $\sigma^*(N)$ is critical for large enough $N$.

\begin{figure}[!ht]
  \centering  
  \includegraphics[width=0.5\columnwidth]{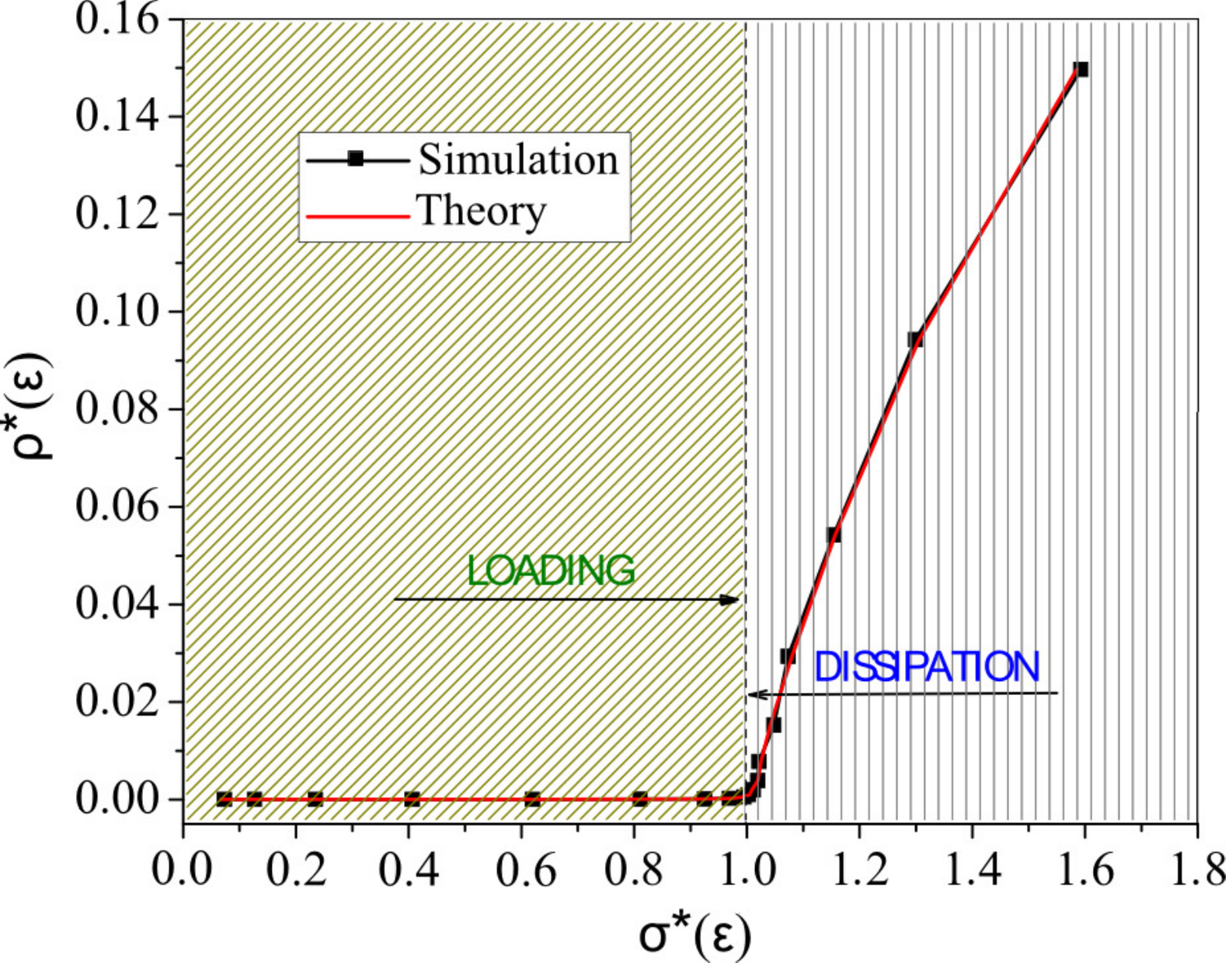}
  \caption{Density of active sites is null for a subcritical system ($\sigma<1$) 
and grows for the supercritical systems. Theoretical curve (mean field) 
and simulation points for the annealed model.  Arrow to the right indicates the effect of synaptic recovery
(mediated by parameters $\epsilon$ and $A$); arrow to the left indicates the effect of synaptic depression, 
induced by avalanches and mediated by the $u$ parameter.
Simulations of supercritical systems are obtained by using large values of $\epsilon$, 
see Fig.~\ref{sig_vs_eps_teo+sim}.
Parameters are: $u = 0.1$, $A = 0.9$, $K = 10$, $n = 3$ and $N = 10000$.}
\label{rho_vs_sig}
\end{figure}

In Fig.~\ref{rho_vs_sig} we present theoretical (Eq.~(\ref{rho_fcao_sig})) 
and simulation results for the annealed model. To show the supercritical regime,
we used large values for $\epsilon$ given a finite $N$ in order to produce
$\rho^*(N) > 0$. From this figure it is clear that
the synaptic dynamics induces the system to lie at the critical point of
an absorbing continuous phase transition~\cite{Munoz99, Dickman00}. 
This is an important feature, not present in the LHG model, 
as extensively discussed by Bonachela and Mu\~noz~\cite{Bonachela10}.

\begin{figure}[ht!]
\centering
\includegraphics[width=0.5\columnwidth]{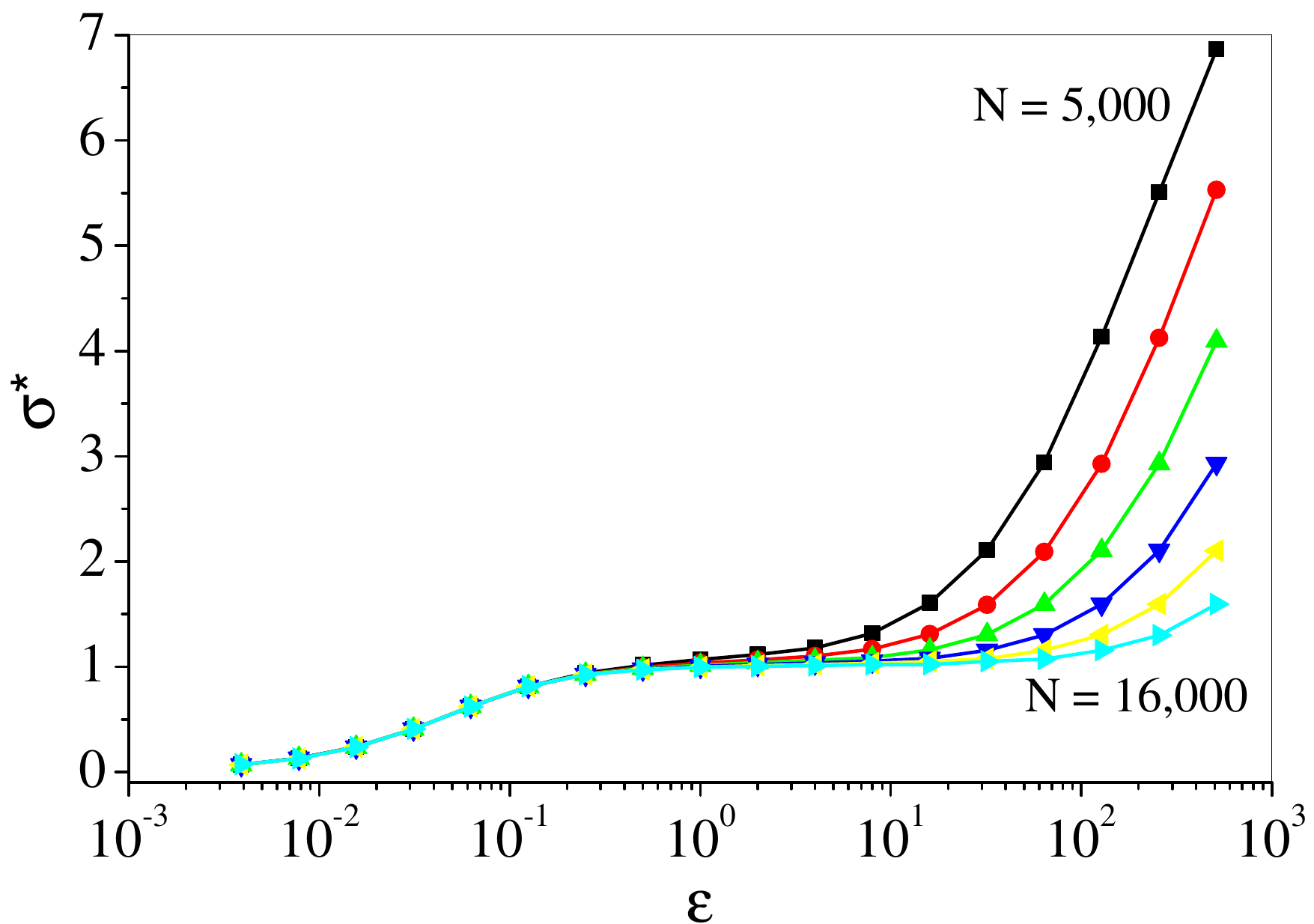}
\caption{Dependence of the stationary branching ratio $\sigma^*$ 
as a function of $\epsilon$ with $N = 500$, $1000$, $2000$, $4000$, $8000$ and $16000$. 
Symbols are obtained from simulations and lines are mean field results. 
Parameters as in Fig.~\ref{rho_vs_sig}.}
\label{sig_vs_eps_teo+sim}
\end{figure}

The main characteristic of our model is that if we plot $\sigma^*(N,p)$ versus a parameter $p$ (for example,
$\epsilon$, $A$ or $u$), not only $\sigma^*(N,p)$ tends to $\sigma_c$, but also 
the parametric derivative $d\sigma^*(N,p)/dp$ tends to zero as $N$ increases. This means 
that a plateau appears around the critical point, so that the parametric dependence 
(for all parameters) vanishes for large $N$. This can be seen explicitly in the mean field equations, where, for
any parameter $p$, we have:
\begin{equation}
\frac{d\sigma^*(p)}{dp} \approx C_p / N,
\end{equation}
for some $p>p_{min}$.
For example, the emergence of a parametric plateau for $\epsilon$ can 
be seen in Fig.~\ref{sig_vs_eps_teo+sim} (notice the logarithmic scale for $\epsilon$). 
The same behavior can be observed for parameters $A$ and $u$. 

\begin{figure}[!ht]
 \centering  
 \includegraphics[width=0.5\columnwidth]{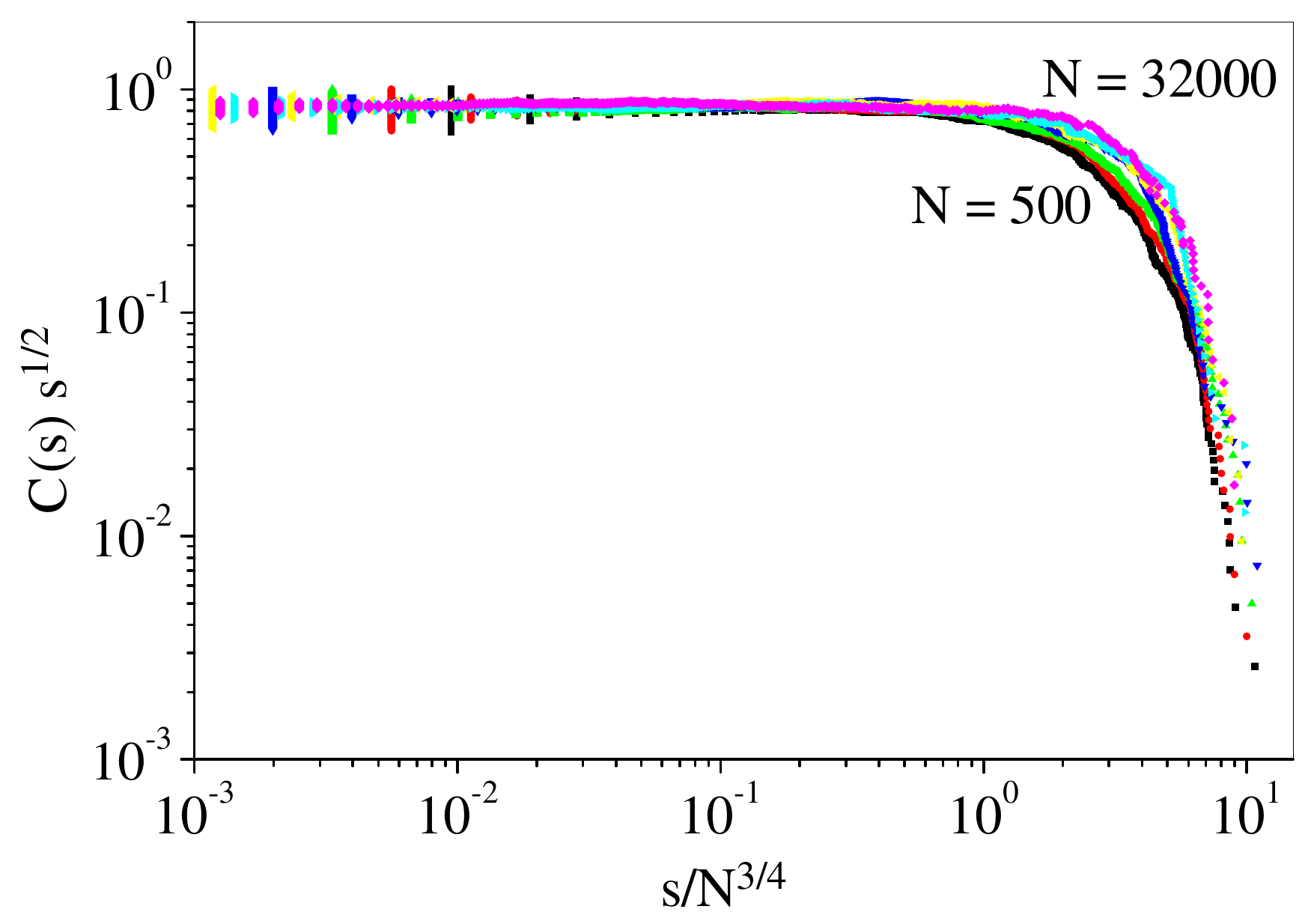}
 \caption{Data collapse of avalanche size ($s$) cumulative distributions $C(s)$ with the scaled parameter  
 $\epsilon_c=0.05N^{1/3}$ for different network sizes, with $n=3$, $u=0.1$, $A=1.0$ and
 $K=10$. The cutoff exponent is $3/4$.}
\label{ava_N}
\end{figure}

The avalanche finite-size scaling, however, is somewhat problematic, as also observed in other non-conservative 
models~\cite{Bonachela09, Bonachela10}.
To obtain a precise scaling of critical avalanches for finite $N$, one
needs to tune the parameters. For example, with other
parameters fixed, the choice 
\begin{equation}
\label{epsilon}
\epsilon = \epsilon_c N^{1/3}\:\:\:,
\end{equation} 
as suggested by Bonachela {\it et al.} \cite{Bonachela09,Bonachela10} 
leads to the correct scaling of the cumulative avalanche size distribution: $C(s)s^{1/2} = f(s/s_0(N))$,
where  $C(s) \equiv \int_s^\infty P_{size}(s) ds$ and $P_{size}(s)$ 
is the probability that an avalanche has size $s$ (see Fig.~\ref{ava_N}).

The scaling with $N$ in Eq.~\ref{epsilon} is not so problematic, since it can be absorbed in the original scaling of the
synaptic recovery, Eq.~\ref{Pdyn}, that is, by using from start a scaling $\epsilon/KN^{2/3}(A-P_{ij})$. 
However, critical avalanches are observed only for a definite choice of $\epsilon_c(A,u)$  (which now does not depend on $N$).
To which extent this tuning implies that our system is a SOqC model is discussed below.

In our model, the cutoff scales as $s_0 \sim N^{a}$ with $a = 3/4$, 
which is different from the scaling found in the LHG model ($a=1$)~\cite{Bonachela10} or other models 
($a=2/3$)~\cite{Bonachela09}. Bonachela {\it et al.}~\cite{Bonachela10} observed that a random neighbor version of the LHG model
presented an anomalous cutoff exponent, but did not reported its value. Naive scaling considerations, similar to that done 
in~\cite{Bonachela09}, although produce $a=2/3$, do not produce $a=3/4$ nor $a=1$, so we prefer to reserve this issue for future
considerations.

\section{Discussion}

Bonachela {\it et al.} \cite{Bonachela09, Bonachela10} tried to define an universality
class, with a definite field theory, for bulk dissipative models that they call Self-Organized quasi-Criticality (SOqC).
In doing so, they claimed that this class is characterized by three (necessary) features:
\begin{itemize}
\item A) The stationary distribution of couplings values $P^*(J)$, which corresponds to 
$P^*(\sigma)$ in our model, has finite variance even in the infinite size limit. The system hoovers around
the critical point, with excursions on the supercritical and subcritical phases. The avalanche distribution
is constructed by summing supercritical and subcritical avalanches;
\item B) The relevant phase transition associated to SOqC is a dynamical percolation transition,
not a continuous absorbing phase transition like conservative SOC models~\cite{Munoz99, Dickman00};
\item C) For finite $N$, to obtain a correct scaling with power law avalanches, we must use tuned
parameters (like $\epsilon_c,A_c,u_c$).
\end{itemize}

The LHG model~\cite{Levina07, Bonachela10} presents all these features,
being classified as a SOqC model. The same occurs with other bulk dissipative models~\cite{Bonachela09}.
Our model, however, presents only feature C (see Fig.~\ref{ava_N}) 
and lacks the important features A and B. Our model, in contrast to the LHG model, 
presents vanishing variance for $P^*(\sigma)$, so that it does not oscillates nor make supercritical (or subcritical)
excursions in the large $N$ limit. Also, the limit $\sigma(N) \rightarrow \sigma_c$ is achieved very fast,
with weak dependence (tuning) on the parameters $A,\epsilon,u$, because they are constants in front of a $1/N$ factor.
Anyway, we observe that, in practice, neuronal networks always work with a very large number
of elements (say, one million), compared, for example, with our simulations with $N=32000$. So,
the large $N$ limit is the relevant one, and in this limit the avalanche behavior depends very weakly on
the parameters, as can be seen from the mean-field results (Eq.~\ref{largeN}).
So, for large $N$, it is more precise to talk about ''gross-tuning" instead of ''fine-tuning" 
to describe the finite size avalanche behavior of our model.

Our model lies at the border of a phase transition to an absorbing state (Compact Directed Percolation), 
instead of a dynamical percolation transition, which is the relevant transition in bona fide 
SOC models~\cite{Dickman00,Munoz99,Bonachela09}.
Since the universality class of our model is different of the LHG model, it should not be
put in the same SOqC class. The problem here is of definition: if item C is {\it sufficient} to classify
a system as SOqC, then our model is SOqC (but them the SOqC will be comprised of two universality classes). 
But if items A and B are {\it necessary} conditions, then
our model is not SOqC and another class must be created.
What we can claim at this moment is that, since our model lacks features A and B, its behavior resembles much more
the conserving SOC models than the LHG model or other non-conservative models~\cite{Bonachela09, Bonachela10}.

The synaptic depression, mediated by the $u$ parameter, is not conservative.
The absence of conserved quantities in the
bulk and specially during the (self-organization) transient is another feature that puts our model apart from conventional SOC models.
The fact our model violates conservation in the bulk, however, is not an impeditive factor for true criticality.
Recently, Moosavi and Montakhab~\cite{Moosavi14} showed that sandpile models with noise (that violates
microscopic conservation but preserves average conservation) can be critical if the noiseless
model is critical and the noise has zero mean. In the case of our model, the conservation on average
is achieved in the stationary state, after a non-conserving transient. So, we conclude
that non-conservative bulk dynamics is not a sufficient feature to put a system in the SOqC class.

Which ingredients could account for the differences between our model and the LHG model, which clearly
pertain to different universality classes?
We identify three main possibilities:
i) their model uses continuous-time integrate-and-fire units 
in contrast to our excitable (SIRS) discrete time units; 
ii) their units are deterministically coupled via weighted synaptic sums, 
while our discrete automata are coupled by probabilistic multiplicative synapses;
iii) in the LHG model, the avalanches are deterministic and, in our model, they are stochastic;
iv) their model is based on a complete 
graph with $N(N-1)$ synapses, while our model sits on a random graph with 
finite average degree $K$ and hence $NK$ synapses. 

It seems to us that items i), ii), iii) and iv) could hardly be responsible for a change of universality class.
On the other hand, item iv) refers to a change of topology, along with a change of time scale
in the synaptic dynamics: LHG uses a change of $1/N$ for synapses that already are of
order $1/N$ (because of the complete graph topology), which means that synapses are strongly perturbed along the time. 
In our random neighbor model, since $K = O(1)$, synapses are $O(1)$ and the synaptic
change of $1/N$ per time step is infinitesimal for large $N$. 

This means that the correction in $P_{ij}$ 
diminishes for increasing $N$, preventing large excursions or oscillations around the $\sigma^*$
point. We believe that this ultra-soft synaptic correction is the missing element,
not contemplated in the literature, that produces a SOC model with vanishing variance.
If this is true, then one can predict that a simulation of the Levina {\it et al.} model with finite 
$K$ random neighbors should fall
in our universality class, that is, presenting vanishing variance for the average coupling
in the thermodynamic limit around a CDP transition.

We finally observe that, although $P^*(\sigma)$ tends to a delta function, the distribution
of local couplings $P^*(P_{ij})$ or, equivalently, of local branching 
ratios $P^*(\sigma_i)$, is not a delta function in the 
$N\rightarrow \infty$ limit. The two facts are not in contradiction because $\sigma=\frac{1}{N}\sum_i^N \sigma_i$ is
an average over $N$ sites and the delta function limit is a large $N$ effect. In other words,
the delta function limit is an effect of the law of large numbers for the average $\sigma$ of the $P^*(\sigma_i)$ distribution
(which continues to have finite variance for large $N$).
This means that the model is nontrivial: 
there is sufficient diversity in the couplings (synapses) to mimic a real biological network.

In conclusion, we have presented an excitable (SIRS) automata model for neural networks
with dynamical synapses which seems to pertain to a new universality class: models with
dissipative bulk dynamics that, due to homeostatic mechanisms, achieve a stationary
conservative (in average) dynamics. In this model, like in conservative SOC models,
the relevant transition pertains to the CPD class.
An evolving ``control'' parameter (the architectural branching ratio) self-organizes to
criticality, and its variance around the critical point vanishes in the thermodynamic limit.

\ack
  We would like to thank Carmem P. C. Prado, Renato Tin\'os, and Afshin Montakhab for discussions. 
  We acknowledge financial support from CAPES, CNPq, PRONEX/FACEPE,
  PRONEM/FACEPE  and CNAIPS-USP.

\section*{References}

\bibliography{copelli}

\end{document}